\documentclass[11pt,a4paper]{article}
\usepackage{graphicx}
\usepackage{amsmath}
\usepackage{bm}

\def\itmb{\begin{itemize}}
\def\itme{\end{itemize}}
\def\enmb{\begin{enumerate}}
\def\enme{\end{enumerate}}
\def\eqnb{\begin{equation}}
\def\eqne{\end{equation}}

\title{Biquaternion Signal Processing for Nonlinear Ultrasonics}

\author{Sadataka Furui$^A$ and Serge Dos Santos$^B$  \\
$^A$ Faculty of Science and Engineering, Teikyo University\\
2-17-12 Toyosatodai, Utsunomiya, 320-0003 Japan {\thanks
{\textit{E-mail address:} furui@umb.teikyo-u.ac.jp}}\\
$^B$ INSA Centre Val de Loire; Université de Tours,\\ INSERM, Imaging Brain \& Neuropsychiatry iBraiN U1253
 F-41034 Blois Cedex, France {\thanks {\textit{E-mail address:} serge.dossantos@insa-cvl.fr}}
}

\begin{document}
\maketitle
\begin{abstract}
Localization and classification of scattered nonlinear ultrasonic signatures in 2 dimensional complex damaged media using Time Reversal based Nonlinear Elastic Wave Spectroscopy (TR-NEWS) approach is extended to 3 dimensional complex damaged media. In (2+1)D, i.e. space 2 dimensional time 1 dimensional spacetime, we used quaternion bases for analyses, while in (3+1)D, we use biquaternion bases.

The optimal weight function of the path of ultrasonic wave in (3+1)D lattice is obtained by using the Echo State Network (ESN) which is a Machine Learning technique. 
The hysteresis effect is incorporated by using the Preisach-Mayergoyz model.
\end{abstract}

 {\bf Keywords:} Biquaternion, Non Destructive Testing, Hysteresis, Nonlinear Elastic Wave Spectroscopy, Noncommutative Geometry
\maketitle

\section{Introduction}
In (2+1)D image processing, Time Reversal based Nonlinear Elastic Wave Spectroscopy (TR-NEWS) was successful which is based on quaternions\cite{SFDS23a,DSCSS06,DSSF16}.  
Details of TR-NEWS experiment of detecting clacks in a block of layered Carbon Fibre Reinforced Polymer (CFRP) and 2D simulation using Newmark algorithm are given in \cite{LSDS17}. From differences of outputs from positive input and from negative inputs of receivers positioned at different positions, the position of a clack in the block was estimated.
In a paper of Fink's group\cite{MRDNF01}, cross-correlation of Time Reversal Mirror (TRM) technique\cite{BLEFF16} was presented. 

In (3+1)D image processing, the mapping by quaternion is replaced by that of biquaternion $M_2({\bf H})$ which is written in the text book of Garling\cite{Garling11}. 
\begin{equation}
j ({\mathcal A}^+_{3,1})=\left(\begin{array}{cc}
a_1 +a_2{\bf k}& b_1{\bf i}+b_2{\bf j}\\
c_1{\bf i}+c_2{\bf j}&d_1+d_2{\bf k}\end{array}\right),\nonumber
\end{equation}
where $a_i,b_i,c_i ,d_i\quad( i=1,2)$ are real and ${\bf i, j, k}$ are the pure quaternion.
The ultrasonic(US) wave is regarded as a phonon propagating in (3+1)D lattices filled with Weyl fermions represented by biquaternions.
We modify the fixed point (FP) action of DeGrand et al.\cite{DGHHN95} used in the simulation of quantum chromo dynamics (QCD) by replacing Dirac fermions to Weyl fermions. The paths in (3+1)D consist of 1) A-type: paths on a 2D plane, 2) B-type: paths on 2 2D parallel planes expanded by ${\bf e}_1, {\bf e}_2$ connected by two links ${\bf e}_1\wedge {\bf e}_2$ and ${\bf e}_2\wedge{\bf e}_1$, 3) C-type: paths expanded by ${\bf e}_1,{\bf e}_2,{\bf e}_3$ and ${\bf e}_4$, which contain time delay or histeresis effects.
Actions of A-type and B-type aree presented in \cite{SFDS23a,SFDS23b}, and we consider in this work, C-type paths

The total length of a path is restricted to be less than or equal to 8 lattice units.  Since we ignore the path that runs along the periphery of a plaquette twice, there are 7 C-type paths. $L19,L20,L21,L22,L23,L24,L25$ in the notation of \cite{DGHHN95}.

Structure of this presentation is as follows. In section 2, we show the paths in (3+1)D spacetime which is optimized by ESN. In section 3, we present how to incorporate time delay or hysteresis effect in lattice simulations.
The Monte Carlo simulation result is shown in section 4. Hysteresis effects and Feynman's path integral are discussed in section 5.
Quaternions appear in the framework of Noncommutative geometry. Noncommutatie geometry and quaternion Fourier transform are reviewed in section 6.
Conclusion and outlook are presented in section 7.

\section{The paths in (3+1)D spacetime}
In the Table 1, direction of paths in the biquaternion basis ${\bf e}_i {\bf e}_j$ and $x,y,z,t$ bases of 8 steps followed by TR 8 steps are shown. In the biquaternion basis, $i$ or $j$ of a step is chosen to be equal to that of the precedent and that of the subsequent step. The time step are ${\bf e}_1{\bf e}_4,{\bf e}_2 {\bf e}_4$ or ${\bf e}_3{\bf e}_4$.  When a step of $t$ or $-t$ are fixed to be ${\bf e}_k{\bf e}_4$, we fix the partner of $t$ or $-t$ to be the same. Except $L21,L22$, $k$ is uniquely fixed.
\begin{table*}[ht]
\begin{center}
\caption{Directions of the wave front of the C-type paths. The first line is in $R^4$ basis, the second line is in biquaternion basis.  }
{\small
\begin{tabular}{r|cccccccccccccccc}
 step & 1&2&3 &4&5&6&7&8&9&10&11&12&13&14&15&16\\
\hline
L19&x&y&z&t&-z&-t&-x&-y&-x&-y&-z&-t&z&t&x&y\\
&23&31&12&24&-12&-24&-23&-31&-23&-31&-12&-24&12&24&23&31\\
\hline
L20&x&y&z&t&-z&-y&-x&-t&-x&-y&-z&-t&z&y&x&t\\
&23&31&12&24&-12&-31&-23&-24&-23&-31&-12&-24&12&31&23&24\\
\hline
L25&x&y&z&t&-x&-y&-z&-t&-x&-y&-z&-t&x&y&z&t\\
&23&31&12&24&-23&-13&-12&-24&-23&-31&-12&-24&23&13&12&24\\
\hline
L21 &x&y&z&t&-z&-x&-t&-y&-x&-y&-z&-t&z&x&t&y\\
 &23&31&12&14/24&-12&-23&-34&-13&-23&-31&-12&-14/24&12&23&34&13\\
\hline
L22&x&y&z&t&-z&-x&-y&-t&-x&-y&-z&-t&z&x&y&t\\
&23&31&12&14/24&-12&-23&-31&-34&-23&-31&-12&-14/24&12&23&31&34\\
\hline
L23&x&y&z&t&-y&-x&-t&-z&-x&-y&-z&-t&y&x&t&z\\
&23&31&12&14&-31&-23&-24&-12&-23&-31&-12&-14&31&23&24&12\\
\hline
L24&x&y&z&t&-y&-x&-z&-t&-x&-y&-z&-t&y&x&z&t\\
&23&31&12&14&-31&-23&-12&-24&-23&-31&-12&-14&31&23&12&24\\
\end{tabular}
}
\end{center}
\end{table*}
In ESN there are recurrent layers of nonlinear units and linear, memory-less read-out layers which are trained. The matrices of input to output $W_{io}$ and input to reservoir $W_{ro}$ are fixed in \cite{Bianchi17}, however in our case, input to output $W_{io}$ is produced from $W_{ro}$.
\begin{figure*}[htb]
\begin{minipage}{0.47\linewidth}
\begin{center}
\includegraphics[width=6cm,angle=0,clip]{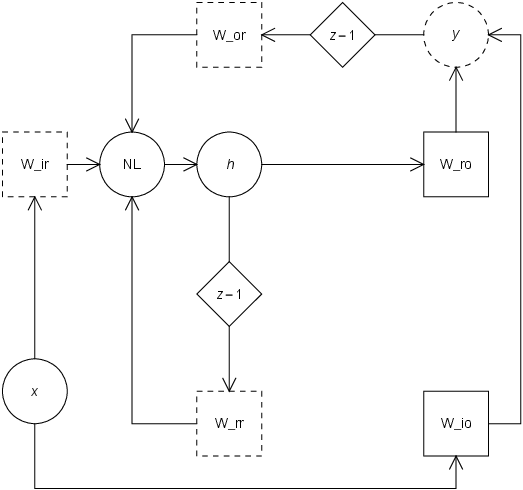}
\end{center}
\end{minipage}
\quad 
\begin{minipage}{0.47\linewidth}
\begin{center}
\includegraphics[width=6cm,angle=0,clip]{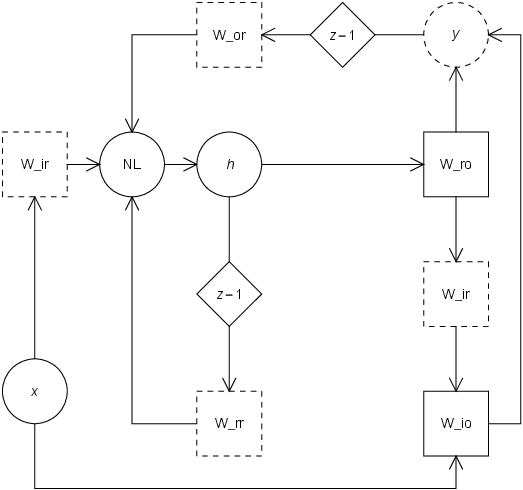}
\end{center}
\end{minipage}
\caption{Standard ESN (left) and our scheme of ESN. $W_{io}$ is produced by a multiplication of $W_{ir}$ on $W_{ro}$.(right) }.\label{ESN2}
\end{figure*}

\begin{figure*}[b]
\begin{minipage}{0.47\linewidth}
\begin{center}
\includegraphics[width=4cm,angle=0,clip]{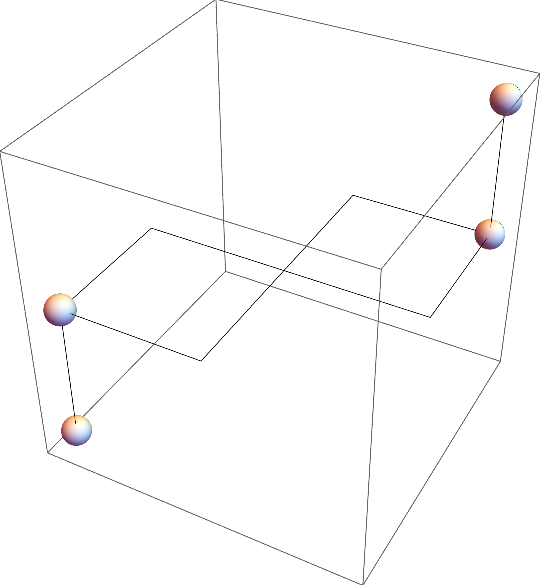}
\end{center}
\end{minipage}
\quad
\begin{minipage}{0.47\linewidth}
\begin{center}
\includegraphics[width=4cm,angle=0,clip]{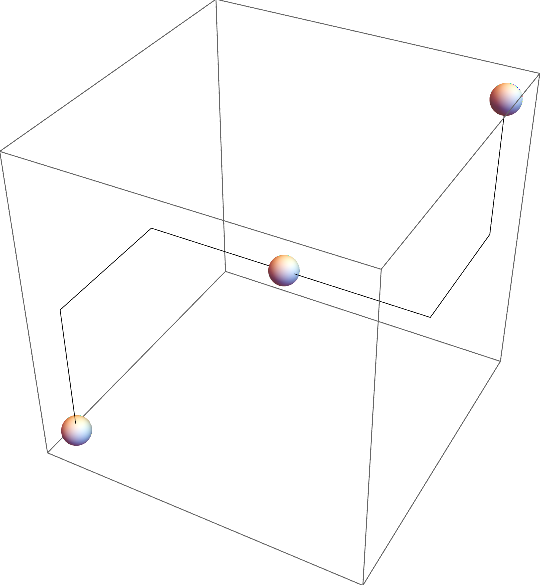}
\end{center}
\end{minipage}
\caption{The path of $L19$(left) and that of $L20$(right). Balls are the places where hysteretic time shift occurs. }\label{l1920}
\end{figure*}

\begin{figure*}[htb]
\begin{minipage}{0.47\linewidth}
\begin{center}
\includegraphics[width=4cm,angle=0,clip]{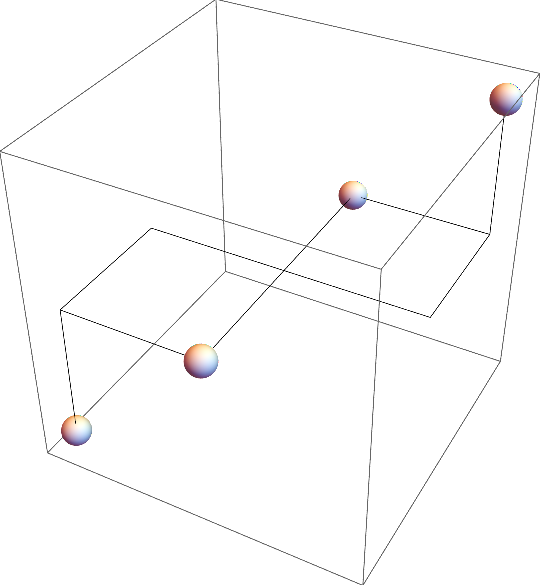}
\end{center}
\end{minipage}
\quad
\begin{minipage}{0.47\linewidth}
\begin{center}
\includegraphics[width=4cm,angle=0,clip]{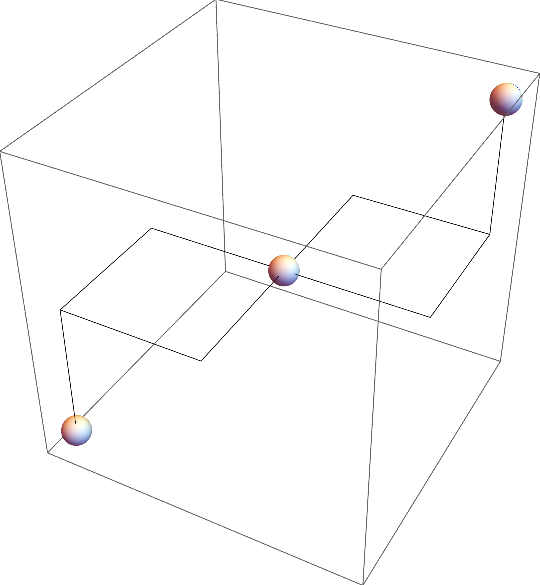}
\end{center}
\end{minipage}
\caption{The path of $L21$(left) and that of  $L22$(right).}\label{l2122}
\end{figure*}

\begin{figure*}[htb]
\begin{minipage}{0.47\linewidth}
\begin{center}
\includegraphics[width=4cm,angle=0,clip]{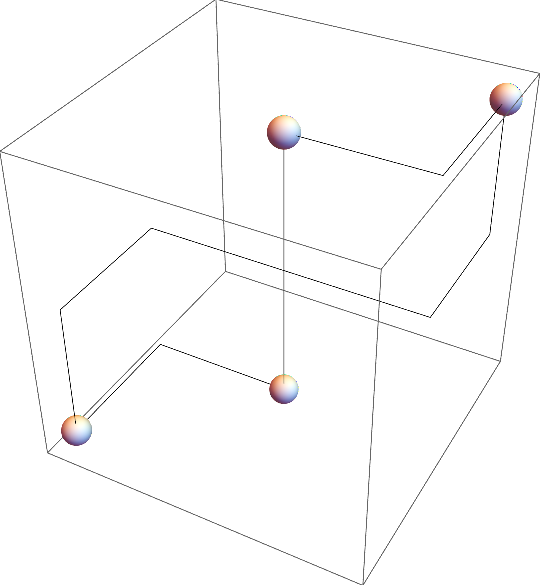}
\end{center}
\end{minipage}
\quad
\begin{minipage}{0.47\linewidth}
\begin{center}
\includegraphics[width=4cm,angle=0,clip]{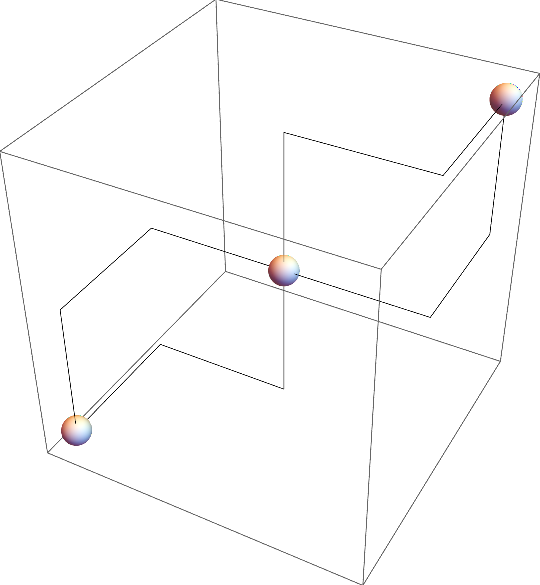}
\end{center}
\end{minipage}
\caption{The path of $L23$(left) and that of $L24$(right).}\label{l2324}
\end{figure*}

\begin{figure}[htb]
\begin{center}
\includegraphics[width=4cm,angle=0,clip]{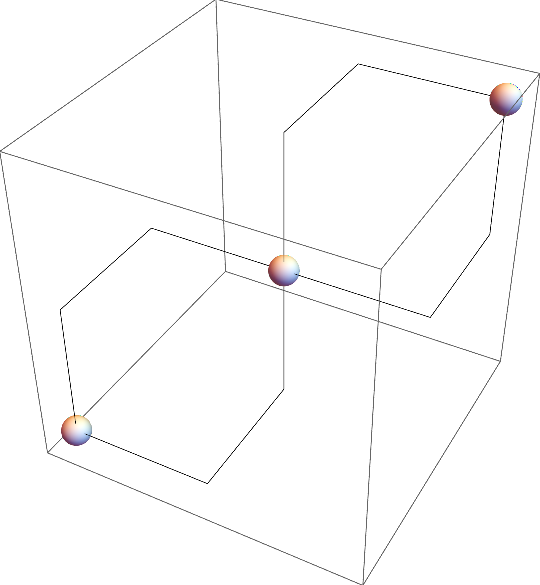}
\end{center}
\caption{The path of $L25$.}\label{l25}
\end{figure}

The 16 steps of $L19,L20$, $L21,L22$, $L23,L24$ and $L25$ are shown in Fig.\ref{l1920} , Fig. \ref{l2122}, Fig.\ref{l2324} and Fig.,\ref{l25}, respectively.  
At balls, time shifts occur. We assume same hysteretic effects occur stochastically in the balls.

The path of $L19, L21, L22$ differ by the points where time shifts  occur, We selected the time shift of $L19$ and $L20$ to be 24 and consider an average of 14 and 24 for $L21$ and $L22$. 

\section{Optimization of the weight function by ESN}
Since the path of $L19,L20,L21,L22,L23,L24,L25$ on the (3+1)D lattice are fixed\cite{DGHHN95}, ${\bf x}[t]$ for the seven paths can be fixed, when the 2D plane of the initial path is selected. Although ${\bf x}[t]$ runs in the 4D space, $W_{ir}{\bf x}[t]$ is a 7 dimensional vector defined on the path at time $t$.

We prepare random matrices
\begin{itemize}
\item  $W_{ro}$ $7\times 6$ matrix, Matrix elements obtained from the FP action..
\item  $W_{io}$ $4\times 6$ matrix. Matrix elements are fixed from $W_{ro}$ and $W_{ir}$.
\item  $W_{ir}$ $4\times 7$ matrix  Matrix elements are subject to be trained.
\item  $W_{or}$ $6\times 7$ matrix. Matrix elements are subject to be trained.
\item  $W_{rr}^{(t)}$ $7\times 7$ matrix. Matrix elements are subject to be trained. Mixing of paths at $t=4,7,8,12,15,16$ have specific filters.
\end{itemize}

By the criterion that at $t=4,7,8,12,15,16$, the paths which don't have the time shift are not disturbed, we multiply a filter to $W_{rr}$ such that non disturbed matrix elements are set to 0.  The interactions between reservoirs at these times are given in the following form.    
\begin{equation} 
{\bf W}_{rr}^{(4)}=\left(\begin{array}{ccccccc}
 *&*&*&*&0&0&*\\
 *&*&*&*&0&0&*\\
 *&*&*&*&0&0&*\\
 *&*&*&*&0&0&*\\
 0&0&*&*&*&*&0\\
 0&0&*&*&*&*&*\\
 *&*&*&*&0&*&*\end{array}\right),\quad
 {\bf W}_{rr}^{(7)}=\left(\begin{array}{ccccccc}
 0&0&0&0&0&0&0\\
 0&0&0&0&0&0&0\\
 0&0&*&0&*&0&0\\
 0&0&0&0&0&0&0\\
 0&0&*&0&*&0&0\\
 0&0&0&0&0&0&0\\
 0&0&0&0&0&0&0\end{array}\right)
\end{equation}
\begin{equation} 
{\bf W}_{rr}^{(8)}=\left(\begin{array}{ccccccc}
 0&0&0&0&0&0&*\\
 0&*&0&*&0&0&*\\
 0&0&0&0&0&0&*\\
 0&*&0&*&0&0&*\\
 0&0&0&0&0&0&0\\
 0&0&0&0&0&*&*\\
 *&*&*&*&0&*&*\end{array}\right),\quad
{\bf W}_{rr}^{(12)}=\left(\begin{array}{ccccccc}
 *&*&*&*&0&0&*\\
 *&*&*&*&0&0&*\\
 *&*&*&*&*&*&*\\
 *&*&*&*&*&*&*\\
 0&0&*&*&*&*&0\\
 0&0&*&*&*&*&*\\
 *&*&*&*&0&*&*\end{array}\right)
\end{equation}
\begin{equation} 
{\bf W}_{rr}^{(15)}=\left(\begin{array}{ccccccc}
 0&0&0&0&0&0&0\\
 0&0&0&0&0&0&0\\
 0&0&*&0&*&0&0\\
 0&0&0&0&0&0&0\\
 0&0&*&0&*&0&0\\
 0&0&0&0&0&0&0\\
 0&0&0&0&0&0&0\end{array}\right),\quad
{\bf W}_{rr}^{(16)}=\left(\begin{array}{ccccccc}
 0&0&0&0&0&0&0\\
 0&*&0&*&0&*&*\\
 0&0&0&0&0&0&0\\
 0&*&0&*&0&0&0\\
 0&0&0&0&0&0&0\\
 0&*&0&0&0&*&*\\
 0&*&0&0&0&*&*\end{array}\right)
\end{equation}

The low and the column are in the order $L19,L20,L21,L22,L23,L24,L25$ and $*$ indicate time shifts in the own path occur or time shifts that cause mixing of paths occur. 

We prepare 40 random  vectors for $t=4 :{\bf h}[4]=f(W_{ir}{\bf x}[4])=( {h}_1[4], {h}_2[4],\cdots, {h}_7[4])$ corresponding to $L19,L20,\cdots , L25$. Here $f$ is the nonlinear activation function, ${\bf x}[4]=$\\ $({x}_1[4],{x}_2[4],{x}_3[4],{x}_4[4])$, and $W_{ir}({x}_1[4],{x}_2[4],{x}_3[4],{x}_4[4])$ is the seven dimentional vector given by the fixed point action of DeGrand et al.\cite{DGHHN95}.  The matrix $W_{io}$ is normalized such that $\sum_{i=1}^4 \sum_{j=1}^6 (W_{io})_{i,j}^2=\sum _{r=1}^7\sum_{j=1}^6 (W_{ro})_{r,j}^2$.

We define the loss function ${\mathcal L}=S_{ro}^2$, where $S_{ro}=W_{ro}{\bf h}-{\bf y}^*$. In the calculation of ${\bf h}[t]$, we choose
\begin{equation}
{\bf h}[t]=f(W_{io}{\bf x}[t]+W_{ro}({\bf h}[t]+{\bf b}_h))
\end{equation}
where 
\begin{equation}
{\bf b}_h=-2\eta\frac{\partial {\mathcal L}}{\partial S_{ro}}\frac{\partial S_{ro}}{\partial h}=-2\eta(W_{ro}{\bf h}-{\bf y}^*)W_{ro}.
\end{equation}

The weight matrix of reservoirs ${W}_{rr}^{(t)}$ can also be trained to yield proper outputs. We assume that the matrix element of ${W}_{rr}^{(t)}$, denoted as $W_{m,n}^{(t)}$ is not 0, when the direction of the path $m$: $e_i e_4$ and that of $n$ : $e_j e_4$ coincide. At $t=4,7,8,12,15$ and 16, there are time shift points between different paths.   Time shifts occur with biquaternion base 14,24 or 34, and when the shift of different paths have the same biquaternion base, we assume that mixings of paths occur and the matrix element differs from 0. 

Optimiztion steps for $W_{rr}^{(t)}$ are done by choosing as an activation function the sigmoid function
 $\frac{1}{1+exp(-W_{rr}^{(t)}{\bf h}[t]-W_{ir}{\bf x}[t+1]-W_{or}{\bf y}[t])}$
which has the range $(0,1]$ and its derivative with respect to $W_{rr}^{(t)}$ as
\begin{equation}
\frac{1-\frac{1}{1+exp(-W_{rr}^{(t)}{\bf h}[t]-W_{ir}{\bf x}[t+1]-W_{or}{\bf y}[t])}}{1+exp(-W_{rr}^{(t)}{\bf h}[t]-W_{ir}{\bf x}[t+1]-W_{or}{\bf y}[t])}
{\bf h}[t]=D W_{rr}^{(t)}{\bf h}[t],
\end{equation} 
or the $\tanh$ functon $\tanh(-W_{rr}^{(t)}{\bf h}[t]-W_{ir}{\bf x}[t+1]-W_{or}{\bf y}[t])$ which has the range $[-1,1]$ and its derivative with respect to $W_{rr}^{(t)}$ is
\begin{equation} 
-{\rm sech}^2(-W_{rr}^{(t)}{\bf h}[t]-W_{ir}{\bf x}[t+1]-W_{or}{\bf y}[t]){\bf h}[t]=DW_{rr}^{(t)}{\bf h}[t].
\end{equation}
$W_{rr}^{(t)}$ is modified to
\begin{equation}
W_{rr}^{(t)}-\eta\frac{\partial{\mathcal L}}{\partial W_{rr}^{(t)}}=W_{rr}^{(t)}-\eta D W_{rr}^{(t)}.
\end{equation}
The optimization of $W_{rr}^{(t)}$ and the state vector ${\bf h}[t]$ are done as follows.
\begin{itemize}
\item Using the ${\bf h}[4]=f(W_{ir}{\bf x}[4])$, where ${\bf x}[4]$ is produced randomly, we calculate ${\bf y}[4]=g(W_{io}{\bf x}[4]+W_{ro}{\bf h}[4])=({y}_1[4],\cdots, {y}_6[4])$, where $g$ and $f$ are the logistic sigmoid function or $\tanh$ function. 

From the second cycle, calculate ${\bf h}[4]=f(W_{rr}^{(4)}{\bf h}[3]+W_{ir}{\bf x}[4]+W_{or}{\bf y}[3])$ and ${\bf y}[4]=g(W_{io}{\bf x}[4]+W_{ro}{\bf h}[4]).$ Modify $W_{rr}^{(4)}\to W_{rr}^{(4)}-\eta D W_{rr}^{(4)}$.

\item Calculate ${\bf h}[5]=f(W_{rr}^{(5)}{\bf h}[4]+W_{i r}{\bf x}[5]+W_{or}{\bf y}[4])=({h}_1[5], {h}_2[5],\cdots,{h}_7[5])$ and ${\bf y}[5]=g(W_{io}{\bf x}[5]+W_{ro}{\bf h}[5])$. Modify $W_{rr}^{(5)}\to W_{rr}^{(5)}-\eta D W_{rr}^{(5)}$.

\item Calculate ${\bf h}[6]=f(W_{rr}^{(6)}{\bf h}[5]+W_{i r}{\bf x}[6]+W_{or}{\bf y}[5])$ and ${\bf y}[6]=g(W_{io}{\bf x}[6]+W_{ro}{\bf h}[6])$. 

Modify $W_{rr}^{(6)}\to W_{rr}^{(6)}-\eta D W_{rr}^{(6)}$.
\item Calculate ${\bf h}[7]=f(W_{rr}^{(7)}{\bf h}[6]+W_{i r}{\bf x}[7]+W_{or}{\bf y}[6])$ and ${\bf y}[7]=g(W_{io}{\bf x}[7]+W_{ro}{\bf h}[7])$.

Modify $W_{rr}^{(7)}\to W_{rr}^{(7)}-\eta D W_{rr}^{(7)}$.

\item Calculate ${\bf h}[8]=f(W_{rr}^{(8)}{\bf h}[7]+W_{i r}{\bf x}[8]+W_{or}{\bf y}[7])$ and ${\bf y}[8]=g(W_{ro}{\bf h}[8])$.

Modify $W_{rr}^{(8)}\to W_{rr}^{(8)}-\eta D W_{rr}^{(8)}$.

\item Calculate ${\bf h}[9]=f(W_{rr}^{(9)}{\bf h}[8]+W_{i r}{\bf x}[9]+W_{or}{\bf y}[8])$ and ${\bf y}[9]=g(W_{io}{\bf x}[9]+W_{ro}{\bf h}[9])$.

Modify $W_{rr}^{(9)}\to W_{rr}^{(9)}-\eta D W_{rr}^{(9)}$.
\item Calculate ${\bf h}[10]=f(W_{rr}^{(10)}{\bf h}[9]+W_{i r}{\bf x}[10]+W_{or}{\bf y}[9])$ and ${\bf y}[10]=g(W_{io}{\bf x}[10]+W_{ro}{\bf h}[10])$. Modify $W_{rr}^{(10)}\to W_{rr}^{(10)}-\eta D W_{rr}^{(10)}$.

\item Calculate ${\bf h}[11]=f(W_{rr}^{(11)}{\bf h}[10]+W_{i r}{\bf x}[11]+W_{or}{\bf y}[10])$ and ${\bf y}[11]=g(W_{io}{\bf x}[11]+W_{ro}{\bf h}[11])$.
Modify $W_{rr}^{(11)}\to W_{rr}^{(11)}-\eta D W_{rr}^{(11)}$.
\item Calculate ${\bf h}[12]=f(W_{rr}^{(12)}{\bf h}[11]+W_{i r}{\bf x}[12]+W_{or}{\bf y}[11])$ and ${\bf y}[12]=g(W_{io}{\bf x}[12]+W_{ro}{\bf h}[12])$. 

Modify $W_{rr}^{(12)}\to W_{rr}^{(12)}-\eta D W_{rr}^{(12)}$.
 
\item Calculate ${\bf h}[13]=f(W_{rr}^{(13)}{\bf h}[12]+W_{i r}{\bf x}[13]+W_{or}{\bf y}[12])$ and ${\bf y}[13]=g(W_{io}{\bf x}[13]+W_{ro}{\bf h}[13])$.
Modify $W_{rr}^{(13)}\to W_{rr}^{(13)}-\eta D W_{rr}^{(13)}$.
\item Calculate ${\bf h}[14]=f(W_{rr}{\bf h}[13]+W_{i r}{\bf x}[14]+W_{or}{\bf y}[13])$ and ${\bf y}[14]=g(W_{io}{\bf x}[14]+W_{ro}{\bf h}[14])$.
Modify $W_{rr}^{(14)}\to W_{rr}^{(14)}-\eta D W_{rr}^{(14)}$.
\item Calculate ${\bf h}[15]=f(W_{rr}^{(15)}{\bf h}[14]+W_{i r}{\bf x}[15]+W_{or}{\bf y}[14])$ and ${\bf y}[15]=g(W_{io}{\bf x}[15]+W_{ro}{\bf h}[15])$.
Modify $W_{rr}^{(15)}\to W_{rr}^{(15)}-\eta D W_{rr}^{(15)}$.

\item Calculate ${\bf h}[16]=f(W_{rr}^{(16)}{\bf h}[15]+W_{i r}{\bf x}[16]+W_{or}{\bf y}[15])$ and ${\bf y}[16]=g(W_{ro}{\bf h}[16])$.

Modify $W_{rr}^{(16)}\to W_{rr}^{(16)}-\eta D W_{rr}^{(16)}$ 

\item Calculate ${\bf h}[1]=f(W_{rr}^{(1)}{\bf h}[16]+W_{i r}{\bf x}[1]+W_{or}{\bf y}[16])$ and ${\bf y}[1]=g(W_{io}{\bf x}[1]+W_{ro}{\bf h}[1])$.

Modify $W_{rr}^{(1)}\to W_{rr}^{(1)}-\eta D W_{rr}^{(12)}$

\item Calculate ${\bf h}[2]=f(W_{rr}^{(2)}{\bf h}[1]+W_{i r}{\bf x}[2]+W_{or}{\bf y}[1])$ and ${\bf y}[2]=g(W_{io}{\bf x}[2]+W_{ro}{\bf h}[2])$.

Modify $W_{rr}^{(2)}\to W_{rr}^{(2)}-\eta D W_{rr}^{(2)}$.

\item Calculate ${\bf h}[3]=f(W_{rr}^{(3)}{\bf h}[2]+W_{i r}{\bf x}[3]+W_{or}{\bf y}[2])$ and ${\bf y}[3]=g(W_{io}{\bf x}[3]+W_{ro}{\bf h}[3])$.
 
Modify $W_{rr}^{(3)}\to W_{rr}^{(3)}-\eta D W_{rr}^{(3)}$.
\end{itemize}
The calculation of ${\bf h}[4]$ changes from the first cycle to ${\bf h}[4']=f(W_{rr}{\bf h}[3]+W_{i r}{\bf x}[4]+W_{or}{\bf y}[3])$.
The cycle from 4 to 20 continues until optimized $W_{ir}, W_{rr}, W_{or}$ are found.

Store ${\bf S}=\left[\begin{array}{cc}
{\bf x}^T[4]& {\bf h}^T[4]\\
{\bf x}^T[7]& {\bf h}^T[7]\\
{\bf x}^T[8]& {\bf h}^T[8]\\
{\bf x}^T[12]& {\bf h}^T[12]\\
{\bf x}^T[15]& {\bf h}^T[15]\\
{\bf x}^T[16]&{\bf h}^T[16]
\end{array}\right]$ and ${\bf y}^*=\left[\begin{array}{c}
{\bf y}^*[4]\\
{\bf y}^*[7]\\
{\bf y}^*[8]\\
{\bf y}^*[12]\\
{\bf y}^*[15]\\
{\bf y}^*[16]
\end{array}\right]$.

As in the case of $(2+1)D$, we optimize the weight function of 7 paths ($L19,L20,L21,L22,L23, L24, L25$) in $(3+1)D$ that minimize the loss ${\mathcal L}=||{\bf S}\,{\bf W}-{\bf y}^*||^2$, where ${\bf W}=[W_{io}, W_{ro}]^T$.
\begin{equation}
{\bf S}{\bf W}=\left[\begin{array}{cc}
(W_{io}{\bf x})^T[4]&(W_{ro}{\bf h})^T[4]\\
(W_{io}{\bf x})^T[7]&(W_{ro}{\bf h})^T[7]\\
(W_{io}{\bf x})^T[8]&(W_{ro}{\bf h})^T[8]\\
(W_{io}{\bf x})^T[12]&(W_{ro}{\bf h})^T[12]\\
(W_{io}{\bf x})^T[15]&(W_{ro}{\bf h})^T[15]\\
(W_{io}{\bf x})^T[16]&(W_{ro}{\bf h})^T[16]
\end{array}\right].
\end{equation}
We adopt a cylindrical lattice model, such that 7 paths start from the origin of a space and returns to the origin. The total action becomes 0 when the path returns to the origin. Therefore at $t=8$ and $t=16$,  $W_{ir}{\bf x}[t]$ become 0 at these epochs.

We are considering paths in momentum space, the Fourier transform of the path in the position space. In ${\bf x}[t]$ of the calculation we consider the projection space $RP^4$ which means that we optimize the scale of the 4D vectors.

We can distinguish the path of time shift through ${e}_1 {e}_4$ and through ${e}_2 {e}_4$, and consider 9 dimensional bases and perform the Elman Recurrent Neural Network (ERNN)\cite{Bianchi17,BSURS15}. 

\section{Monte Carlo simulation results}
The searched output ${\bf y}[t]$ of ESN, which is 7 dimensional vector, and $t$ runs from $t=4,\cdots 16, 1, 2, 3$ modulus 16. We want to obtain weights of 7 bases which are stable over many cycles. We observed that for the nonlinear activation function, $\tanh$ is better than the sigmoid function. 

The $\tanh$ function was used in the Long Short Term Memory (LSTM) cells method\cite{RLM22}.

The deviations $||{\bf y}[t]-{\bf y}^*[t]||^2$ using the $\tanh$ function are much reduced from those obtained by the sigmoid function. However, near $t=9$ and $t=16$, large deviation appear. It is due to the fact that at $t=8$ and $t=16$, the random walk returns to the original position, and  $y^*[t]$ is taken to be zero. The good property of this method is that from the 3000th cycle to the 4000th cycle, the weight is stable. After 5000 cycles, the absolute value of the correlation increases slghtly. 

The obtained ${\bf y}[9]$ and ${\bf y}[16]$ suggests that ${\bf y}^*[8]={\bf y}^*[16]=0$ derived from the action of paths returning to the original in 3D space meight be inappropriate. 

\begin{figure*}[htb]
\begin{minipage}{0.47\linewidth}
\begin{center}
\includegraphics[width=6cm,angle=0,clip]{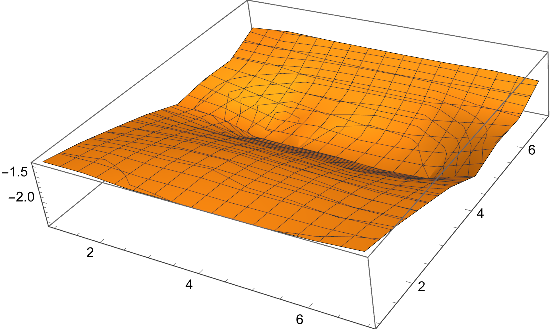}
\end{center}
\end{minipage}
\quad
\begin{minipage}{0.47\linewidth}
\begin{center}
\includegraphics[width=6cm,angle=0,clip]{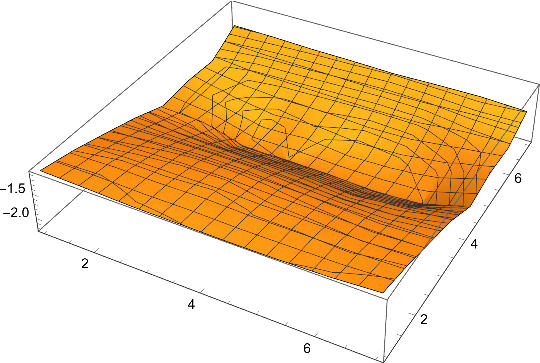}
\end{center}
\end{minipage}
\caption{The correlation function of $W_{ro}{\bf h}$ and $\bf y$ at the 3000th cycle (left), at the 4000th cycle (right),  }\label{gCor}
\end{figure*}

 In order to visualize the stability of outputs, we calcurated the correlation of $H=W_{ro}{\bf h}$ and $Y={\bf y}$ both $7\times 7$ matrices
 which is expressed as
 \begin{table*}[htb]
 \begin{equation}
 \left(\begin{array}{ccccccc}
\frac{\sigma_{H_1Y_1}}{\sigma_{H_1}\sigma_{Y_1}}&\frac{\sigma_{H_1Y_2}}{\sigma_{H_1}\sigma_{Y_2}}&\frac{\sigma_{H_1Y_3}}{\sigma_{H_1}\sigma_{Y_3}}&\frac{\sigma_{H_1Y_4}}{\sigma_{H_1}\sigma_{Y_4}}&\frac{\sigma_{H_1Y_5}}{\sigma_{H_1}\sigma_{Y_5}}&\frac{\sigma_{H_1Y_6}}{\sigma_{H_1}\sigma_{Y_6}}& \frac{\sigma_{H_1 Y_7}}{\sigma_{H_1}\sigma_{Y_7}}\\
\vdots&\vdots&\vdots&\vdots&\vdots&\vdots&\vdots\\
\frac{\sigma_{H_7 Y_1}}{\sigma_{H_7}\sigma_{Y_1}}&\frac{\sigma_{H_7 Y_2}}{\sigma_{H_7}\sigma_{Y_2}}&\frac{\sigma_{H_7 Y_3}}{\sigma_{H_7}\sigma_{Y_3}}&\frac{\sigma_{H_7 Y_4}}{\sigma_{H_7}\sigma_{Y_4}}&\frac{\sigma_{H_7 Y_5}}{\sigma_{H_7}\sigma_{Y_5}}&\frac{\sigma_{H_7 Y_6}}{\sigma_{H_7}\sigma_{Y_6}}&\frac{\sigma_{H_7,y_7}}{\sigma_{H_7}\sigma_{Y_7}}\end{array}\right)
 \end{equation}
 \end{table*}
 From the 3000th cycle to 4000th cycle,  $\frac{\sigma_{H_i Y_j}}{\sigma_{H_i}\sigma_{Y_j}}$ multiplied by 100 are plotted in Figure\ref{gCor}.
Correlations at the 3000th cycle and at the 4000th cycle are similar. As the number of cycles increases, the depth of correlation surface becomes deeper.

We tried to run the program up to 10000 cycles, but the Mathematica produced a warning that the accuracy may be decreased. Therefore, we consider data between 3000 cycles and 4000 cycles and abort our choice of ${\bf y}^*[t]$, and  took ${\bf y}^*[t]=0$ for all $t$.

\begin{figure*}[htb]
\begin{minipage}{0.47\linewidth}
\begin{center}
\includegraphics[width=8cm,angle=0,clip]{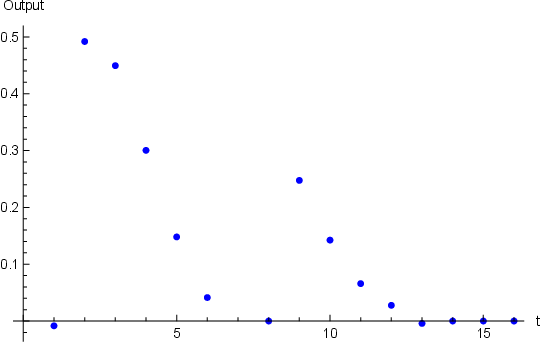}
\end{center}
\end{minipage}
\quad
\begin{minipage}{0.47\linewidth}
\begin{center}
\includegraphics[width=8cm,angle=0,clip]{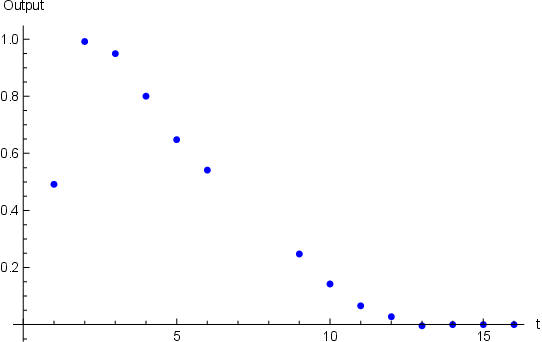}
\end{center}
\end{minipage}
\caption{Output calculated by the weight function at the 4000th cycle and the Fixed Point action at $t=2,\cdots, 16$ (left).
The output in the range $0<t<8$ shifted upward by 0.5. (right)}\label{output}
\end{figure*}

The obtained weight function is multiplied by the fixed point action at $t=T/16, T=0,\cdots, 255$ which is used in the previous calculation\cite{SFDS23b}. Actions of $L21$ and $L22$ are chosen to be the average of $e_1e_4$ and $e_2e_4$ contributions. The actions at $t=8$ and $t=16$ are 0, and the output in the range $9\leq t\leq 12$ and $4\leq t\leq 6$ are almost parallel.

We shifted the output in the range $0<t<8$ of all cycles by adding 0.5, and found that outputs in $0<t<8$ and $8<t<16$ become smooth. 
The shift of action in $0<t<8$ can be interpleted as an additional action of Preisach-Meyergoyz (PM) model\cite{Mayergoyz03} which is discussed in the next section.

\section{Hysteresis effects and Feynman's path integral}
Feynman showed the probability amplitude for a space-time path in the region $R$ as $|\varphi(R)|^2$\cite{Feynman48}, where
\begin{equation}
\varphi(R)=\lim_{\epsilon\to 0}\int_R exp[\frac{\sqrt{-1}}{\hbar}\sum_i S(x_{i+1},x_i)]\cdots\frac{dx_{i+1}}{A}\frac{dx_i}{A}\cdots.
\end{equation}
Here $\frac{1}{A}$ is a normalization factor to make the wave function satisfies the Schr\"odinger equation. The probability that the path is in $R'$ and later in $R''$ is $|\varphi(R',R'')|^2$, where 
\begin{eqnarray}
&&\varphi(R',R'')=\int \chi^*(x,t)\psi(x,t) dx\nonumber\\
&&\psi(x_k,t)=\lim_{\epsilon\to 0}\int_{R'}exp[\frac{\sqrt{-1}}{\hbar}\sum_{t=-\infty}^{k-1}S(x_{i+1},x_i)]\frac{dx_{k-1}}{A}\frac{d x_{k-2}}{A}\cdots,\nonumber\\
&&\chi^*(x_k,t)=\lim_{\epsilon\to 0} \int_{R''} exp[\frac{\sqrt{-1}}{\hbar}\sum_{i=k}^\infty S(x_{i+1}, x_i)]\frac{dx_{k+1}}{A}\frac{dx_{k+2}}{A}\cdots.
\end{eqnarray}

The wave function at $t=t+\epsilon$ is approximated as
\begin{equation}
\psi(x_{k+1},t+\epsilon)=\int exp[\frac{\sqrt{-1}}{\hbar}S(x_{k+1}, x_k)]\psi(x_k,t) \frac{dx_k}{A}.
\end{equation}

The PM model\cite{Mayergoyz03} contains many building blocks of $\varphi(R', R'')$ with time delay. It is formulated by the input $u(t)$ and output $f(t)$ as\cite{PKDS12}
\begin{equation}
f(t)=\hat \gamma u(t)=\int\int_{\alpha\geq \beta}\mu(\alpha,\beta)\hat\gamma_{\alpha\beta} u(t) d\alpha d\beta,
\end{equation}
where 
\begin{equation}
\hat\gamma_{\alpha\beta} u(t)=\left\{ \begin{array}{cc}
1,& u(t)\geq \alpha,\\
0,& u(t)\leq \beta,\\
k,& u(t)\in(\beta,\alpha),\end{array}\right.
\end{equation}
\begin{equation}
k=\left\{\begin{array}{cc}
1&\exists t^*:u(t^*)>\alpha \quad {\rm and}\quad \forall\tau\in (t^*,t), u(\tau)\in (\beta,\alpha),\\
0&\exists t^*:u(t^*)<\beta \quad {\rm and}\quad \forall\tau\in (t^*,t),u(\tau)\in (\beta,\alpha).\end{array}\right.
\end{equation}
\begin{figure*}[htb]
\begin{minipage}{0.47\linewidth}
\begin{center}
\includegraphics[width=3cm,angle=0,clip]{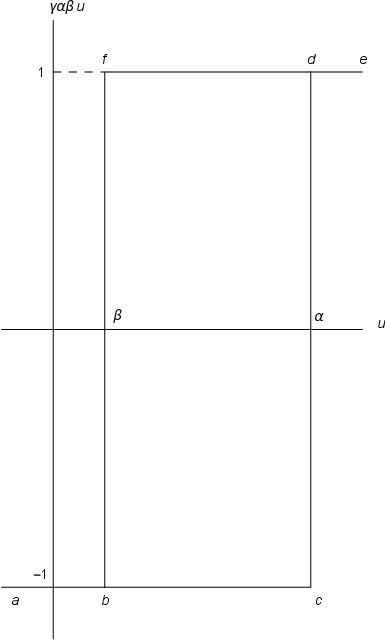}
\end{center}
\end{minipage}
\quad
\begin{minipage}{0.47\linewidth}
\begin{center}
\includegraphics[width=4.5cm,angle=0,clip]{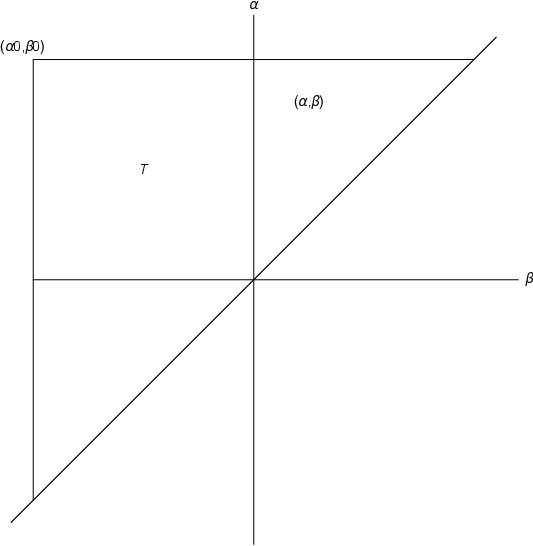}
\end{center}
\end{minipage}
\caption{The $\hat \gamma_{\alpha\beta} u(t)$(left) and the integral area on the $(\beta,\alpha)$ plane(right). }\label{preisach1}
\end{figure*}
\begin{figure*}
\begin{minipage}{0.47\linewidth}
\begin{center}
\includegraphics[width=6cm,angle=0,clip]{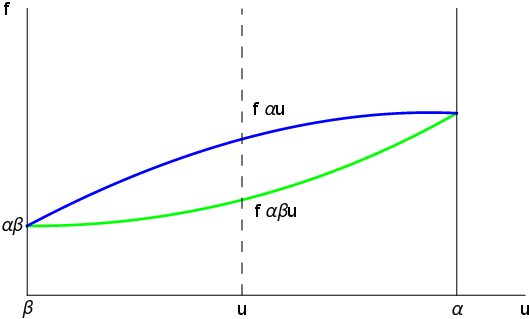}
\end{center}
\end{minipage}
\hfill
\begin{minipage}{0.47\linewidth}
\begin{center}
\includegraphics[width=6cm,angle=0,clip]{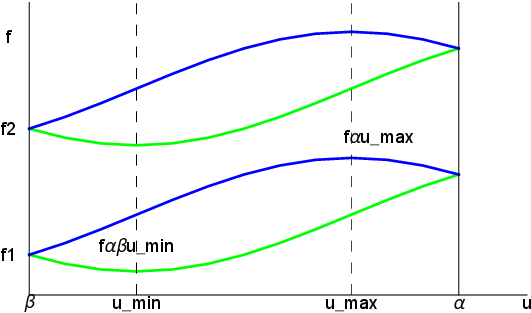}
\end{center}
\end{minipage}
\caption{\small $f_{23}$ (green) and $f_{32}$ (blue) as a function of input $u$. Left is the Preisach-Mayergoyz model. Right is quaternion basis model. }
\end{figure*}

\begin{figure}[htb]
\begin{center}
\includegraphics[width=8cm,angle=0,clip]{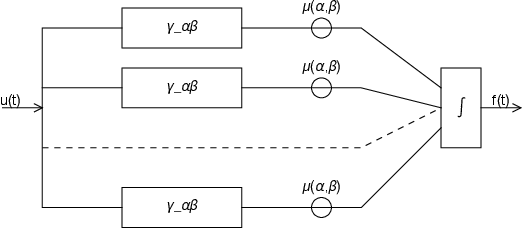}
\end{center}
\caption{Input $u(t)$ and output $f(t)$ of the Preisach-Meyergoyz Model.}\label{PMGraph}
\end{figure}

The PM model is constructed as a superposition of $\gamma_{\alpha\beta}u(t)$ multipied by the weight function $\mu(\alpha,\beta)$,
as shown in the Fig.\ref{PMGraph}, given in \cite{Mayergoyz03}.

In magnetism, input $u(t)$ corresponds to the magnetic field $H(t)$, and $\gamma_{\alpha\beta}u(t)$ corresponds to the magnetization $M(t)$.

Hysteresis effects calculated by using quaternion basis differ from those of standard PM model. 
We also remark that the 4D echo technique in medical examinations is a record of 3D images in the time series, and hysteresis effects are not considered.

\section{Noncommutative geometry and quaternion Fourier transform}
 Quaternion and biquaternion basis model can be used not only for NDT, but also for QCD lattice simulations. Quaternion quantam Mechanics was proposed by Finkelstein et al.\cite{FJSS62} in 1962, and Adler\cite{Adler85,Adler94} used quaternions in generalized quantum dynmics, and in operator gauge invariant quaternionic field theory. He considered the total trace Lagrangean and Hamiltonian dynamics and asked, "Given two scalar or fermion quaternionic operator fields, is there a criterion for determining whether they are related by a bi-unitary operator gauge transformation?"\cite{Adler94}.
Our systems are not related by a gauge transformation but related by a different choice of bases.

Quantum mechanics represented by quaternions is proposed also by Connes\cite{Connes94}. 

In Heisenberg picture, the equation of motion is
\begin{equation}
\sqrt{-1}\hbar \partial_t\psi({\bf x},t)=[\psi({\bf x},t),H].
\end{equation}
When there are hysteresis effect, Connes extended the equation of motion using groupoids in dynamical systems expressed by $(X,R,\pi)$,
where $X$ is a topological space, $R$ is the real number space, and $\pi$ is a mapping from $X\times R$ to $X$. 

Algebraic structures of a Groupoid $G$ and its distinguished subset $G^{(0)}$ are characterised by the source map $s$ and the range map $r$ on an element $\gamma\in G$
\begin{equation}
(a*b)(\gamma)=\sum_{\gamma_1\circ \gamma_2=\gamma} a(\gamma_1)b(\gamma_2)
\end{equation}
where $a,b$ are arbitrary maps and
\begin{equation}
\circ: G^{(2)}=\{(\gamma_1,\gamma_2)\in G\times G; s(\gamma_1)=r(\gamma_2)\}\to G.
\end{equation}
The source map and the range map satisfy
\begin{enumerate}
\item $s(\gamma_1\circ\gamma_2)=s(\gamma_2), r(\gamma_1\circ \gamma_2)=r(\gamma_1)$.
\item $s(x)=r(x)=x, x\in G\to G^{(0)}$.
\item $\gamma\circ s(\gamma)=\gamma$,  $r(\gamma)\circ \gamma=\gamma$.
\item $(\gamma_1\circ\gamma_2)\circ \gamma_3=\gamma_1\circ(\gamma_2\circ\gamma_3)$.
\item $\gamma \gamma^{-1}=r(\gamma)$, $\gamma^{-1}\gamma=s(\gamma)$.
\end{enumerate}
We consider $G^{(0)}\subset M\times [0,1]$ with inclusion
\begin{eqnarray}
&&({\bf x},\epsilon)\to ({\bf x,x},\epsilon)\in M\times M\times [0,1]\quad {\rm for}\quad x\in M,\epsilon>0 \nonumber\\
&&({\bf x},0)\to {\bf x}\in M\subset TM,
\end{eqnarray}
where $TM$ is the tangent manifold defined by the sequence $(x_n,y_n,\epsilon_n)$ in $G_1=M\times M\times ]0,1]$ in the limit of
\begin{equation}
{\bf x}_n\to {\bf x},\quad {\bf y}_n\to {\bf y},\quad \frac{{\bf x}_n-{\bf y}_n}{\epsilon_n}\to X.
\end{equation}

The range map and the source map satisfy
\begin{equation}
\left\{\begin{array}{l}
r({\bf x,y},\epsilon)=(x,\epsilon) \quad {\rm for }\quad x\in M,\epsilon>0\nonumber\\
r({\bf x},X)=({\bf x},0)\quad {\rm for}\quad {\bf x}\in M, X\in T_x(M) \end{array}\right.
\end{equation}
\begin{equation}  
\left\{ \begin{array}{l}
s({\bf x,y},\epsilon)=(y,\epsilon) \quad {\rm for }\quad y\in M\epsilon>0\nonumber\\
s({\bf x},X)=({\bf x},0)\quad {\rm for}\quad {\bf x}\in M, X\in T_x(M). \end{array}\right.
\end{equation}
The composition is
\begin{eqnarray}
&&({\bf x,y},\epsilon)\circ({\bf y,z},\epsilon)=({\bf x,z},\epsilon) \quad{\rm for}\quad \epsilon>0\nonumber\\
&& \quad{\rm and}\quad {\bf x,y,z}\in M,\nonumber\\
&&({\bf x},X)\circ ({\bf x},Y)=({\bf x},X+Y)\quad {\rm for}\quad {\bf x}\in M\quad\nonumber\\
&&\quad{\rm and}\quad X,Y\in T_x(M).
\end{eqnarray}
\begin{figure}[htb]
\begin{center}
\includegraphics[width=6cm,angle=0,clip]{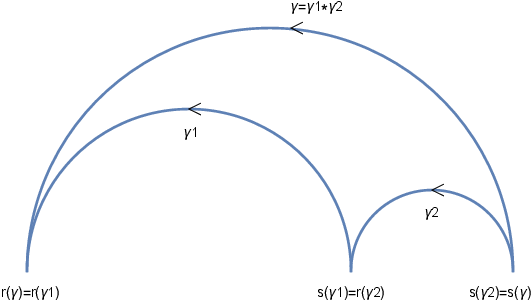}
\end{center}
\caption{The groupoid structure of the hysteresis effect. $r(\gamma)=({\bf x},0)$.}\label{Groupoid}
\end{figure}
To allow hysteresis effects we allow in the case of $\gamma=\gamma_1\circ\gamma_2$, $r(\gamma_2)=({\bf x},1)=s(\gamma_1)$, where 1 is in the unit of time shift in our FP lattice model, and $\bf x$ is the position of the ball in the Figure 3,4 and 5.
The PM model says that the output becomes a sum of Groupoids including time shifts
In the book of \cite{BP93} formulae of measurling cross-correlations in various setups are given. 
For input signal $x(t)$ and output signal $y(t)$ one uses the following quantities
\begin{itemize}
\item $x(t)=$ on going natural input (unmeasured).
\item $i(t)=$ known external input signal (measured)
\item $v(t)=$ linear output (unmeasured) caused by $x(t)$.
\item $r(t)=$ linear output (unmeasured) caused by $i(t)$.
\item $n(t)=$ unknown output noise (unmeasured).
\item $y(t)=v(t)+r(t)+n(t)=$ total output signal (measured).
\end{itemize}
Their Fourier transforms are related as
\begin{eqnarray}
&&R(f,T)=H(f)I(f,T), \quad V(f,T)=H(f)X(f,T)\nonumber\\
&&Y(f,T)=R(f,T)+V(f,T)+N(f,T)\nonumber\\
&&\quad \quad \quad=H(f)[I(f,T)+X(f,T)]+N(f,T)
\end{eqnarray}
where $H(f)$ is the frequency response function, and $T$ is the maximal time. 

The cross-spectrum terms $G_{ix}(f)=G_{in}(f)=G_{nv}(f)=G_{nr}(f)$ are assumed to be 0.
\begin{eqnarray}
&&G_{yy}(f)=G_{rr}(f)+G_{vv}(f)+G_{nn}(f)\nonumber\\
&&G_{iy}(f)=H(f)G_{ii}(f)
\end{eqnarray}
When the external excitation signal is a white noise $G_{ii}(f)=K$ and $H(f)=\frac{G_{iy}(f)}{K}$.

Assume that paths $k=1,\cdots, k=r$ have uniform gain factor, and
\begin{eqnarray}
y(t)&=&H_1 x(t-\tau_1)+H_2(t-\tau_2)+\cdots +H_r x(t-\tau_r)\nonumber\\
&=&\sum_{k=1}^r H_k x(t-\tau_k).
\end{eqnarray} 
The cross-correlation is
\begin{eqnarray}
R_{xy}(\tau)&=&\lim_{T\to\infty}\frac{1}{T}\int_0^T x(t)[H_1 x(t-\tau_1+\tau)+\cdots +H_k x(t-\tau_r+\tau)]d\tau\nonumber\\
&=&\sum_{k=1}^r H_k R_{xx}(\tau-\tau_k).
\end{eqnarray}

For $x(t)=X\sin(2\pi f_0 t+\theta)$, autocorrelation function is
\begin{eqnarray}
R_{xx}(\tau)&=&\frac{X^2}{2\pi}\int_0^{2\pi}\sin(2\pi f_0t+\theta)\sin(2\pi f_0(t+\tau)+\theta)\nonumber\\
&&d\theta=\frac{X^2}{2}\cos(2\pi f_0\tau).
\end{eqnarray}

Autocorrelation of uniform band width 1,
$
G_{xx}(f)=\left\{\begin{array}{c}
G\quad {\rm for }\quad f\leq 1\\
0\quad {\rm for}\quad f>1\nonumber\end{array}
\right.$ is
\begin{equation}
R_{xx}(2(\tau-25)/15)=\int_0^1\cos(2\pi f\tau) df=G \frac{\sin(2\pi \tau)}{2\pi\tau}=G\frac{\sin(\pi\tau)\cos(\pi\tau)}{\pi\tau}.
\end{equation}
\begin{figure*}[htb]
\begin{minipage}{0.47\linewidth}
\begin{center}
\includegraphics[width=6cm,angle=0,clip]{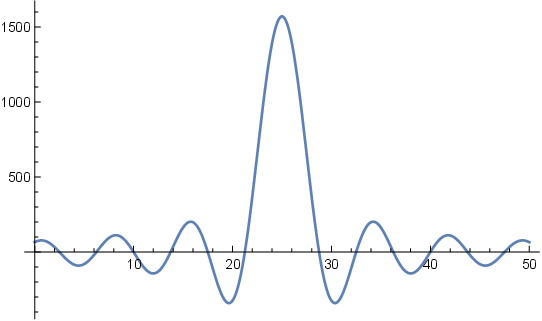}
\end{center}
\end{minipage}
\quad
\begin{minipage}{0.47\linewidth}
\begin{center}
\includegraphics[width=6cm,angle=0,clip]{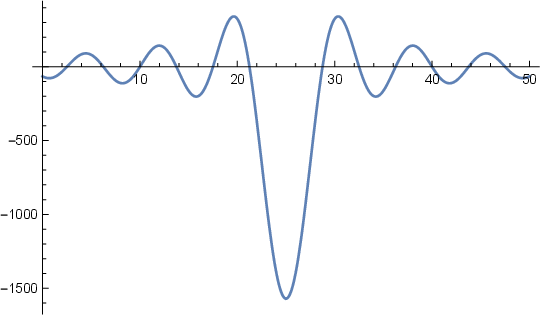}
\end{center}
\end{minipage}
\caption{The autocorrelation of free incident and its TR signals with time delay $\tau=25$. }\label{CC25}
\end{figure*}
Cross-correlation is given by the autocorrelation of $x(t)$ multiplied by $H$ and displayed in time
\begin{eqnarray}
&&R_{xy}(\tau)=\lim_{T\to\infty}\frac{1}{T}\int_0^Tx(t)[Hx(t-\frac{d}{c}+\tau)+n(t+\tau)]dt\nonumber\\
&&=HR_{xx}(\tau-\frac{d}{c}).
\end{eqnarray}

The experiment was taken 64000 epochs, and shows the absolute value of the peak or dip of the side-lobes is about 1000 three times larger than the free wave correlations. Enhancement of $R_{xy}^{25}(2(\tau-25)/15)$ near $\tau=17.5$ and 32.5 can be attributed to the nonlinear effects.

The cross-correlation function is the Fourier transform of the spectral function 
\begin{equation}
H(f)=\frac{G_{xy}(f)}{G_{xx}(f)}=\frac{S_{xy}(f)}{S_{xx}(f)}
\end{equation}
\begin{equation}
R_{xy}(\tau)=\int_{-\infty}^\infty S_{xy}(f)e^{\sqrt {-1}2\pi f\tau} df.
\end{equation}
For two single sample records $x(t)$ and $y(t)$, unbiased estimate of cross-correlation is
\begin{equation}
\hat R_{xy}(\tau)=\frac{1}{T-\tau}\int_0^{T-\tau} x(t)y(t+\tau) dt.
\end{equation}
When there are multiple paths correlation coefficient associated with $\rho_{xy}(d_k/c_k)$, where $d_k$ is the length of the path and $c_k$ is the sound velocity, and $\sum_{k=1}^r \rho_{xy}^2(\frac{d_k}{c_k})=1$.

The random error in $\hat R_{xy}(\tau)$ is, when data is a collection of $n_d$ records and $N$ cycles,
\begin{equation}
\epsilon[\hat R_{xy}(\tau)]=[\frac{1+\rho_{xy}^{-2}(\tau)} {N n_d}]^{1/2}.
\end{equation}

We show our analysis using two inputs many outputs case, using the Focused Decomposition of Time Reversal Operator [D\'ecomposition de l'Operateur Retournement Temporel (DORT)] (FDORT) method\cite{BLEFF16,RBCBF06}.  Two inputs are incident and time reversed wave \cite{DSSF16} and the forcusing in TR Acoustics (TRA) is a method of concentrating acoustic energy in  a small volume with  the diameter close to a half wavelength\cite{DSCSS06}. 

The optimal weight function of the C-type fixed point actions which contain hysteresis effect can be simulated by using the biquaternion basis. We observed stablility of the action on the output layer produced from the reservoirs. A main difference from the ERNN in our application is that there exists output from the reservoir and mixing among reservoir $W_{rr}$ exist in ESN, while in ERNN, in our application output from additional 2 inputs exist and no direct output from the hidden layer. ERNN in general contains outputs from hidden layers, but in our ERNN, which is based on the Convolutional Neural Network (CNN), outputs had direct connection with values in the input layers.  The ESN has more freedom of incorporating hysteresis effects. In ESN, we obtained the optimal weight function and multiplying the fixed point action and taking the average of $e_1e_4$ and $e_2e_4$ contributions and combining $0<t<8$ part and $8<t<16$ part, we obtained  smooth $t$ dependent action output. 

For establishing our methods, comparison between experiments and ESN simulation results in lower dimensional system would be useful. Our method can be used for a comparison of data of TR-NEWS experiment in (2+1)D using quaternion bases which are Clifford algebra bases. We analyzed ultrasonic wave propagation with hysteresis effects using biquaternions.
The experiment using the NOVA university sample\cite{KSFLMSOG24} is done by using 3 memristor array to produce TR signals

Excitation Symmetry Analysis Method (ESAM) \cite{DSP08} and FDORT techniques\cite{VPDS09} are applied t obtain amplitudes of four eigenvectors E, A, B1 and B2. In this case $P_{1,2}$ is replaced by $P_E, P_A, P_{B1}$ and $P_{B1}$.

In signal processing, generalization of Fourier transform, Laplace transform, etc. which is called the linear canonical transformation (LCT)
is utilized.  Its extension to quaternion Fourier transform and discussion on Heisenberg's uncertainty princiople are discussed in \cite{BA16,PK23}. In these works, output of quaternion linear time varying system
$R(\xi,b)=\int_{R^2} H_1(\xi,b,t)f(t) dt$ is discussed. The $\xi$ is a vector in $R^2$, and $b$ represents the time shift. 
An extension of quaternion windowed linear canonical transform (QWLCT) to biquaternion system is left for the future.
\section{Conclusion and outlook}
We showed that the weight function of paths defined by the fixed point action can be optimized by the ESN using the $\tanh$ function for expressing the nonlinearity. 


For getting the optimal solution of these problems, Machine Learning techniques can be applied. Nonlinearity and hysteresis could be explored in these basis. As shown in \cite{DSP10}, the technique is applicable for an  extension of dental investigation, as an example, which is restricted at present in (2+1)D system. The damaged position can be detected in the 3D space if receivers and transducers of TR waves are distributed in the 3D space and signals traveling to all directions can be detected. 


\vskip 0.5 truecm

{\bf Acknowledgments}:
SF thanks the Japan Industrial Science Laboratory (Nissanken) for the financial aid of the travel expense to INSA Centre Val de Loire, Blois Campus in November 2023, and Prof. M. Arai and Prof. K. Hamada for allowing the use of workstations in their laboratory.


\begin{thebibliography}{300}
\bibitem{SFDS23a}
S. Furui and  S. Dos Santos, \emph{Application of Quaternion Neural Network to Time Reversal Based Nonlinear Elastic Wave Spectroscopy}, INAE, {\bf 8} 183-199, (2023).%
\bibitem{DSCSS06}
S. Dos Santos, B.K. Choi, A. Sutin and A. Sarvazyan, \emph{Nonlinear imaging Based on Time Reversal Acaustic Forcussing},  CFA 2006 (2006).
\bibitem{DSSF16}
S. Dos Santos and S. Furui, \emph{A memristor based ultrasonic transducer: the memoducer},
IEEE Ultrasonic Symposium IUS 2016. (2016).%
\bibitem{LSDS17}
M. Lints, A. Salpere and S. Dos Santos, \emph{ Simulation of defects in CFRP and delayed TR-NEWS analysis}, Research Report Mech 320/17 (2017).%
\bibitem{MRDNF01}
G. Montaldo, P. Roux, A. Derode, C. Negreira and M. Fink, \emph{:Generation of very high pressure pulse with 1-bit time reversal in a solid waveguide},J. Acoust. Soc. Am. {\bf 110} (6) 2849-2857 (2001).
\bibitem{BLEFF16}
V. Bacot, M. Labousse, A. Eddi, M. Fink and E. Fort, \emph{Time reversal and holography with spacetime transformations}, Nature Physics, {\bf 12} (10) (2016).%
\bibitem{BP93}
J.S. Bendat and A.G. Piersol, \emph{Engineering Applications of Correlation and Spectral Analysis}, Second Edition,
John Wiley  \& Sons (1993).
\bibitem{SFDS23b} 
S. Furui and S. Dos Santos, \emph{Clifford Fourier Transforms in (2+1)D Lattice Simulations of Soliton  Propagations},  PoS  Lattice 2022 (2023),%
\bibitem{Garling11}
D.J.H. Garling, \emph{ Clifford Algebras: An Introduction}, Cambridge University Press (2011).
\bibitem{DGHHN95}
T. DeGrand, A. Hasenfratz, P. Hasenfratz, and F. Niedermayer, \emph{Non-perturbative tests of the fixed point action for SU(3) gauge theory},Nucl. Phys.{\bf B454} 615-637 (1995): arXiv:9506031[hep-lat].%
\bibitem{Feynman48}
R.P. Feynman,  \emph{Space-Time Approach to Non-Relativistic Quantum Mechanics}, Rev. Mod. Phys. {\bf 20} (2) 367-387 (1948)%
\bibitem{Mayergoyz03} 
I. Mayergoyz, \emph{Mathematical Models of Hysteresis and their Applications}, Elsevier, Amsterdam (2003).
\bibitem{Connes94}
A. Connes, \emph{Noncommutative Geometry}, Academic Press, San Diego (1994).
\bibitem{Bianchi17}
F.M. Bianchi  et al. , \emph{Recurrent Neural Networks for Short-Term Load Forecasting}, Springer Briefs in Computer Science, https://doi.org/10.1007/978-3-319-70338-1
\bibitem{BSURS15}
F.M. Bianchi., S. Scardapane, A. Uncini, A. Rizzi and A. Sadeghian,  \emph{Prediction of telephone calls load using Echo State Network with exogeneous variables}, Neural Networks,Elsevier  {\bf 71}, 204-213 (2015).%
\bibitem{MRTK07}
N. Marwan, M.C. Romano, M. Thiel and J. Kurths,\emph{Recurrence plots for the analysis of complex systems}, Physics Reports {\bf 438} 237-329, Elsevier(2007).
\bibitem{SFDS24b}
S. Furui and S. Dos Santos, \emph{Optimization of Weight Function of $R_{3,1}$ Space Time by the Echo State Network}, Int. Journal of Soft Computing (IJSC) {\bf 15} (3) 39-54 (2024).%
\bibitem{PKDS12}
J. Papouskova, V. Kus and S. Dos Santos, \emph{Preisach-Mayergoyz space model density identification for nonlinear physical systems: "L-2" and "D-divergence" minimization methods}, Proceedings of Meetings on Acoustics, {\bf 16}, 045018 (2012).
\bibitem{DDSKP23}
Z. Dvorakova, S. Dos Santos, V. Kus V and Z. Prevolovsky, \emph{Localization and Classification of scattered nonlinear ultrasonic signatures in bio-mechanical media using time reversal approach}, .J. Acoust. Soc. Am. {\bf 154} (3) pp.1684-1695 (2023)%
\bibitem{GCDSBM07}
T. Goursolle, S. Call\'e, S. Dos Santos and O. Bou Matar, \emph{A two-dimensional pseudospectral model for time reversal and nonlinear elastic wave spectroscopy}, J. Accoust. Soc. Am {\bf  122}(6) pp.3220-3229 (2007).%
\bibitem{DSP08}
S. Dos Santos and C. Plag, \emph{Excitation Symmetry Analysis Method (ESAM) for Calculatiion of Higher Order Nonlenearities}, Int. J. Non-Linear Mech. {\bf 43} 104-119 (2008).%
\bibitem{DSP10}
S. Dos Santos and Z. Prevorovsky, \emph{Ultrasonographie dentaire par techniques TR-NEWS}, 10\`eme Congr\`es Francais d'Acoustique, Lyon (2010) %
\bibitem{VPDS09}
S. Vejvodova, Z. Prevorovsky and S. Dos Santos, \emph{Nonlinear Time eversal Tomography of Structural Defects}  Proceedings of Meetings on Acoustics {\bf 3}, 045093 (2009).
\bibitem{DSSF18}
S. Dos Santos A. Masood, S. Furui and G. Nardoni, \emph{Self-calibration of multiscale hysteresis with memristors in nonlinear time reversal based processes}IEEE BEC 2018. (2018).%
\bibitem{RLM22}
S. Raschka, Y. Liu and V. Mirjalili, \emph{Machine Learning with Pytorch and Scikit-Learn}, Packt, Birmingham (2022).
\bibitem{KSFLMSOG24}
S. Kokare, J. Shen, P.P. Fonseca, J.G. Lopes, C.M. Machado, T.G. Santos, J.P. Oliveira and R. Godina, \emph{Wire arc additive manufacturing of a high-strength low-alloy steel part: evironmental impacts, costs, and mechanical properties} The international Journal of Advanced Manufacturing Technology, {\bf 134}, 453 (2024). %
\bibitem{FJSS62}
D. Finkelstein, J.M. Jauch, S.Schiminovich and D. Speiser, \emph{Foundation of Quaternion Quantum Mechanics}, Journal of Mathematical Physics, {\bf 7} (2) 207-220 (1962).
\bibitem{Adler85}
S.L. Adler,  \emph{Quaternionic Quantum Field Theory}, Phys. Rev. Lett. {\bf 55} (8) 783-786 (1985). Errata {\bf 55} (13) 1430 (1985).
\bibitem{Adler86}
S.L. Adler, \emph{Quaternionic Quantum Field Theory}, Commun. Math. Phys, {\bf 104}, 611-656 (1986).
\bibitem{Adler94}
S.L. Adler, \emph{Generalized quantum dynamics}, Nuclear Physics, {\bf B415}, 195-242 (1994)
\bibitem{BA16}
M. Bahri and R. Ashino, \emph{A Simplified Proof of Uncertainty Principle for Quaternion Linear Canonical Transform}, Hindwari
Abstract and Applied Analysis, Volume 2016, Article I 5879430 (2016).
\bibitem{PK23}
A. Prasad and M. Kundu, \emph{Uncertainty principles and Applications of quaternion windowed linear canonical transform}, Elsevier, Optik-International Journal for Light and Electron Optics, {\bf 272} 170220 (2023)
\bibitem{RBCBF06}
J-L. Robert, M. Boucher, C. Cohen-Bacrie and M. Fink, \emph{Time reversal operator decomposition with focused transmmission and robustness to speckle noise:Application to microcalcification detection}, J. Accoust. Soc. Am. {\bf 119}(6) 3848-3859 (2006).
\end{thebibliography}
\end{document}